\documentclass
[prd,preprint,showpacs,groupedaddress]{revtex4-1}
\usepackage{amsmath}
\usepackage{amsfonts}
\usepackage{amssymb}
\usepackage{geometry}
\usepackage{natbib}
\usepackage[english]{babel}
\usepackage{graphicx}
\usepackage{epstopdf}
\begin{document}

  \title{Extended Holographic R\'{e}nyi Entropy and hyperbolic black hole with scalar hair}

  \author{Dao-Quan Sun} \email {lygsdq@outlook.com}

  \affiliation{School of Physics, Sun Yat-sen University, Guangzhou 510275, P. R. China}

  \date{\today}

  \begin{abstract}

We study the extended thermodynamics of the hyperbolic black hole with scalar hair and obtain the extended holographic R\'{e}nyi entropy of holographic conformal field theories with scalar hair. We analyze the behaviors of the extended holographic R\'{e}nyi entropy in terms of holographic calculations. Moreover, we generalize the capacity of entanglement from the extended  R\'{e}nyi entropy and show that it maps to the heat capacity of the thermal conformal field theories on the hyperbolic space.

  \end{abstract}



  \maketitle

  \section{INTRODUCTION}

 The gauge/gravity~\cite{Maldacena:1997re,PhysRevLett.96.181602} correspondence has provided deep connections between quantum field theories, information theory, and black holes in anti-de Sitter space. The thermodynamics of black holes in anti-de Sitter space began with the pioneering paper of Hawking and Page~\cite{Hawking1983}. According to the AdS/CFT duality, the thermal properties of anti-de Sitter(AdS) black holes can be reinterpreted as those of a conformal field theory; see~\cite{Witten:1998zw} for an example. A quantum system may be described by the density matrix $\rho$ in a pure or mixed state. For a bipartite entanglement, we can divide the total system into a subsystem $A$ and its complement $B$.
  The entanglement entropy $S_{EE}$ for a subsystem $A$ is defined by von Neumann entropy $S_{EE}=-\text{Tr}\rho_{A}\log\rho_{A}$, where $\rho_{A}=\text{Tr}_{B}\rho_{\text{total}}$ is the reduced density matrix of the subsystem $A$ and defined by tracing out with respect to $B$ from the total density matrix of the system. In the gravity dual, Ryu and Takayanagi presented the holographic formula of entanglement entropy~\cite{PhysRevLett.96.181602} known as an entropy-area relation $S_{EE}=\frac{\text{Area}(\gamma_{A})}{4G_{N}}$  where $\gamma_{A}$ is a minimal area surface extending into the bulk. Another much studied measure of entanglement is the R\'{e}nyi entropy which is a one-parameter generalization of entanglement entropy.  It is defined as
  \begin{equation}
\label{eq:renyi}
S_{q}=\frac{1}{1-q}\text{log}[\text{Tr}_{A}(\rho ^{q}_{A})],
\end{equation}
and in the parameter $q\to 1$ limit with the normalization $\text{Tr}_{A}(\rho _{A})=1$, the R\'{e}nyi entropy is reduced to entanglement entropy.
 The R\'{e}nyi entropy has a relation to thermodynamics~\cite{Baez:2011upp}. Let the density matrix represent the thermal density matrix
 $\rho=\text{exp}(-H/T_{0})/Z(T_{0})$, where the partition function $Z(T_{0})=\text{Tr} \text{exp}(-H/T_{0})$, $H$ is an appropriate Hamiltonian and the temperature is $T_{0}$. With the parameter $q=T_{0}/T$ appearing as a ratio of temperatures, the R\'{e}nyi entropy can relate to the familiar concept of the Helmholz free energy $F(T)=-T\text{log}Z(T)$ as
 \begin{equation}
\label{eq:renyi-F}
S_{q}=-\frac{F(T)-F(T_{0})}{T-T_{0}}.
\end{equation}
With the standard thermal entropy $S=-\partial F/\partial T $,  the R\'{e}nyi  entropy can be written as:
 \begin{equation}
\label{eq:renyi-St}
S_{q}=\frac{q}{q-1}\frac{1}{T_{0}} \int_{T_{0}/q}^{T_{0}}S(T)dT.
\end{equation}

 In the papers~\cite{Casini:2011kv,Hung:2011nu,Belin:2013uta}, authors showed that the entanglement entropy of a holographic CFT reduced on a round ball of radius $R$ in Minkowski space is equal to the thermal entropy of a hyperbolically sliced AdS-Schwarzschild black hole. Letting $U$ be the unitary operator acting on the CFT Hilbert space, a conformal map can be constructed, and it relates the reduced density matrix $\rho_{A}$ to a thermal one. The map~\cite{Casini:2011kv,Hung:2011nu}
 \begin{equation}
\label{eq:rhov}
\rho_{A}= U^{\dagger} (\frac{e^{-\beta H}}{Z})U,
\end{equation}
where $H$ is Hamiltonian generating time translations in the hyperbolic spacetime $\mathbb{R\times H}^{d-1}$, and the temperature is $T_{0}=\beta^{-1}=1/(2\pi L_{0})$.

According to the extended framework of black hole thermodynamics, the form of $S_{q}$ given in eq.\eqref{eq:renyi-F} was generalized in the paper~\cite{Johnson:2018bma} by using the first law of thermodynamics $\Delta G=-S\Delta T+V \Delta p$, the extended R\'{e}nyi entropy was written as
 \begin{equation}
\label{eq:sqb1}
S_{q,b}=- \frac{G(p_{0},T_{0})-G(b^{2}p_{0},T_{0}/q)}{\Delta T + V_{0}\Delta p/S_{0}}=\frac{1}{(\Delta T + V_{0}\Delta p/S_{0})}\int_{T_{0}/q}^{T_{0}}S(T)(1+\frac{V}{S}\frac{dp}{dT})dT,
\end{equation}
where $\Delta T=T_{0}-T_{0}/q$, $\Delta p= b^{2}p_{0}-p_{0}$, and $\frac{V_{0}}{S_{0}}=\frac{4G_{N}L_{0}}{d}$. Here, the $b$ is another parameter in addition to the parameter $q$. In the limit $b\to 1$, the $S_{q,b}$ reduces to usual R\'{e}nyi entropy. When the parameter $q$,$b\to 1$ the extended R\'{e}nyi entropy $S_{q,b}$ reduces to the entanglement entropy $S_{EE}$.

The field theory interpretation of $S_{q,b}$ arises from a generalizing the conformal map \eqref{eq:rhov}~\cite{Johnson:2018bma}
\begin{equation}
\label{eq:rhov-pv}
\rho_{A}^{(b)}= U^{\dagger} \left(\frac{e^{-(H+b^{2}p_{0}V_{0})/T_{0}}}{Z(T_{0},p_{0})}\right)U,
\end{equation}
which connects the flat space CFT to the thermal ensemble on $\mathbb{R\times H}^{d-2}$. As a consequence we have
\begin{equation}
\label{eq:sqb-rhov}
S_{q,b}= \frac{1}{[(1-q)-q(d-1)(b^{2}-1)/2]}\text{log}[\text{Tr}(\rho_{A}^{(b)})^{q}].
\end{equation}
This shows that $S_{1,b}$ can be computed by using the replica trick, where the parameter $b$ plays a similar role to $q$. Recently, the extended holographic R\'{e}nyi entropy was generalized to the charged case~\cite{Svesko:2020dfw}.

In this paper, we first study the extended thermodynamics of the hyperbolic black hole with scalar hair by considering the cosmological constant as a thermodynamic variable, and obtain the extended holographic R\'{e}nyi entropy of the hairy black holes. We show that the extended holographic R\'{e}nyi entropy presents a transition at critical parameters when the black hole has a thermodynamic phase transition at a critical temperature. We also holographically compute the inequalities of the extended holographic R\'{e}nyi entropy and the conformal dimension of twist operators in terms of gravitational quantities.
As an important quantum information measurement, the capacity of entanglement~\cite{PhysRevLett.105.080501,DeBoer:2018kvc} has been studied by some researchers~\cite{Nakaguchi:2016zqi,Li:2008kda,PhysRevB.96.205108,Kawabata:2021hac,Kawabata:2021vyo,Nandy:2021hmk,Bhattacharjee:2021jff}. In the paper~\cite{Li:2008kda}, it shows that entanglement entropy and capacity of entanglement are two features of the full entanglement spectrum. In addition, recently, the capacity of entanglement has been used to explore the Hawking radiation~\cite{Kawabata:2021hac,Kawabata:2021vyo}. In this work, we generalize the capacity of entanglement with the extended R\'{e}nyi entropy, and show that the extended capacity of entanglement maps to the heat capacity of the thermal conformal field theories on the hyperbolic space.

The paper is organized as follows. In Section \uppercase\expandafter{\romannumeral2}, we first briefly review the neutral hyperbolic black holes with scalar hair in the $\text{AdS}_{4}$ spacetime and then study extended thermodynamics of the black hole. In Section \uppercase\expandafter{\romannumeral3}, we give the extended holographic R\'{e}nyi entropy of the hairy black holes and analyze the behaviors of $S_{q,b}$. Section \uppercase\expandafter{\romannumeral4}, we holographically compute the conformal dimension of twist operators in terms of gravitational quantities. In Section \uppercase\expandafter{\romannumeral5}, we generalize the capacity of entanglement with the extended R\'{e}nyi entropy. Section \uppercase\expandafter{\romannumeral6} is reserved for conclusions and discussions.

\section{Extended the thermodynamics of hairy black hole}

An Einstein-Maxwell-dilaton(EMD) system consists of gravity, a single gauge field and a dilaton field. It has been widely used in gauge/gravity duality. An example is given by~\cite{Gao:2004tu}, whose special case can be embedded into higher dimensional supergravities~\cite{Cvetic:1999xp}. Now, we consider a class of neutral hyperbolic black holes with scalar~\cite{Ren:2019lgw,Bai:2022obp,Martinez:2004nb}. The action is given by
\begin{equation}
\label{eq:action}
S = \int_{}^{}d^{4}x\sqrt{-g}[R-\frac{1}{2}(\partial \phi)^{2}-V(\phi)],
\end{equation}
where the scalar potential is
\begin{equation}
\label{eq:v-pot}
V(\phi) = -\frac{2}{(1+\alpha^{2})L^{2}}[\alpha^{2}(3\alpha^{2}-1)e^{-\phi/\alpha}+8\alpha^{2}e^{(\alpha-1/\alpha)\phi/2}+(3-\alpha^{2})e^{\alpha \phi}],
\end{equation}
and $\alpha$ is a parameter. When $\alpha$ takes $\alpha=0, 1/\sqrt{3}, 1$ and $\sqrt{3}$ it corresponds to special cases
of STU supergravity. The solution of the system is given by~\cite{Ren:2019lgw}
 \begin{equation}
  \begin{aligned}
  \label{eq:d2s=}
 ds^2&=-f(r)dt^2+\frac{dr^2}{f(r)}+U(r)d\Sigma_{2}^{2} \  , \  \  \    e^{\alpha \phi}=\left(1-\frac{b}{r} \right)^{\frac{2\alpha ^{2}}{1+\alpha ^{2}}},
 \end{aligned}
  \end{equation}
 where
 \begin{equation}
  \begin{aligned}
  \label{eq:fr}
 f(r)&=-\left(1-\frac{b}{r} \right)^{\frac{1-\alpha ^{2}}{1+\alpha ^{2}}}+\frac{r^{2}}{L^{2}}\left(1-\frac{b}{r} \right)^{\frac{2\alpha ^{2}}{1+\alpha ^{2}}}, \  \  \    U(r)=r^{2}\left(1-\frac{b}{r} \right)^{\frac{2\alpha ^{2}}{1+\alpha ^{2}}}.
  \end{aligned}
  \end{equation}
The event horizon $r_{h}$ of the black hole is determined by taking the largest and real root of $f(r)=0$,
\begin{equation}
\label{eq:r+-}
r_{h}=r_{\pm} = \left(\frac{2\pi LT(1+\alpha ^{2})\pm \sqrt{4\pi^{2}L^{2}T^{2}(1+\alpha ^{2})^{2}+(3-\alpha ^{2})(1-3\alpha ^{2})}}{3-\alpha ^{2}} \right)^{\frac{1-3\alpha ^{2}}{1+\alpha ^{2}}}L,
\end{equation}
where we may have one or two real solutions of $r_{h}$.

The relevant thermodynamic quantities are given by~\cite{Ren:2019lgw,Bai:2022obp}:
\begin{equation}
  \begin{aligned}
  \label{eq:thermodynamics-quant}
  T=\frac{f^{'}(r_{h})}{4\pi}=\frac{1}{4\pi(1+\alpha ^{2})L}\left[(3-\alpha ^{2})\left(\frac{r_{h}}{L}\right)^{\frac{1+\alpha ^{2}}{1-3\alpha ^{2}}}-(1-3\alpha ^{2})\left(\frac{r_{h}}{L}\right)^{-\frac{1+\alpha ^{2}}{1-3\alpha ^{2}}}\right], \\
    M=-\frac{V_{\Sigma}}{8\pi G_{N}}\frac{1-\alpha ^{2}}{1+\alpha ^{2}}r_{h}\left[1-\left(\frac{r_{h}}{L}\right)^{\frac{2(1+\alpha ^{2})}{1-3\alpha ^{2}}}\right],\\
  S_{BH}=\frac{V_{\Sigma}}{4 G_{N}}\left(\frac{r_{h}}{L}\right)^{\frac{2(1-\alpha ^{2})}{1-3\alpha ^{2}}}L^{2}.
  \end{aligned}
  \end{equation}
Here $T$ is the Hawking temperature of the black hole, $M$ is the mass of the black hole, $S_{BH}$ is the Bekenstein-Hawking entropy, and $V_{\Sigma}$ is the (regulated) volume of the hyperbolic space $\mathbb{H}^{2}$.
When $\alpha=0$, the black hole is the hyperbolic $\text{AdS}_{4}$ black hole without scalar hair. Then $M=0$ corresponds to a special massless  hyperbolic black hole which has a hyperbolic horizon at $r_{h}=L_{0}$, and the temperature is $T_{0}=1/(2\pi L_{0})$.

In the extended thermodynamics of the black hole~\cite{Kubiznak:2016qmn,Frassino:2022zaz,Ahmed:2023snm,Caceres:2015vsa,Li:2018aax,Kastor:2014dra}, the cosmological constant $\Lambda=-\frac{d(d-1)}{2L^{2}}$ as a thermodynamic variable is dynamical, and the thermodynamical pressure $p$ in the extended phase space is
\begin{equation}
\label{eq:pressure}
p=-\frac{\Lambda}{8\pi G_{N}}.
\end{equation}
The mass of a black hole is interpreted as the enthalpy of spacetime~\cite{Kastor:2009wy}. Namely, $M=H(S,p)=U+pV$. The Smarr relation can be written as $M=2TS-2pV$, and the first law of black hole thermodynamics in an extended phase space reads:
\begin{equation}
\label{eq:first law}
dM=TdS+Vdp.
\end{equation}
The corresponding thermodynamic volume is given by
\begin{equation}
\label{eq:volume}
V=\left(\frac{\partial M}{\partial p}\right)_{S}=\frac{V_{\Sigma} L^{3}}{3}\left[\frac{1}{1+\alpha ^{2}}\left(\frac{r_{h}}{L}\right)^{\frac{3-\alpha ^{2}}{1-3\alpha ^{2}}}+\frac{\alpha ^{2}}{1+\alpha ^{2}} \frac{r_{h}}{L} \right].
\end{equation}

 \begin{figure}[htbp]
\centering
\includegraphics[width=7.6cm]{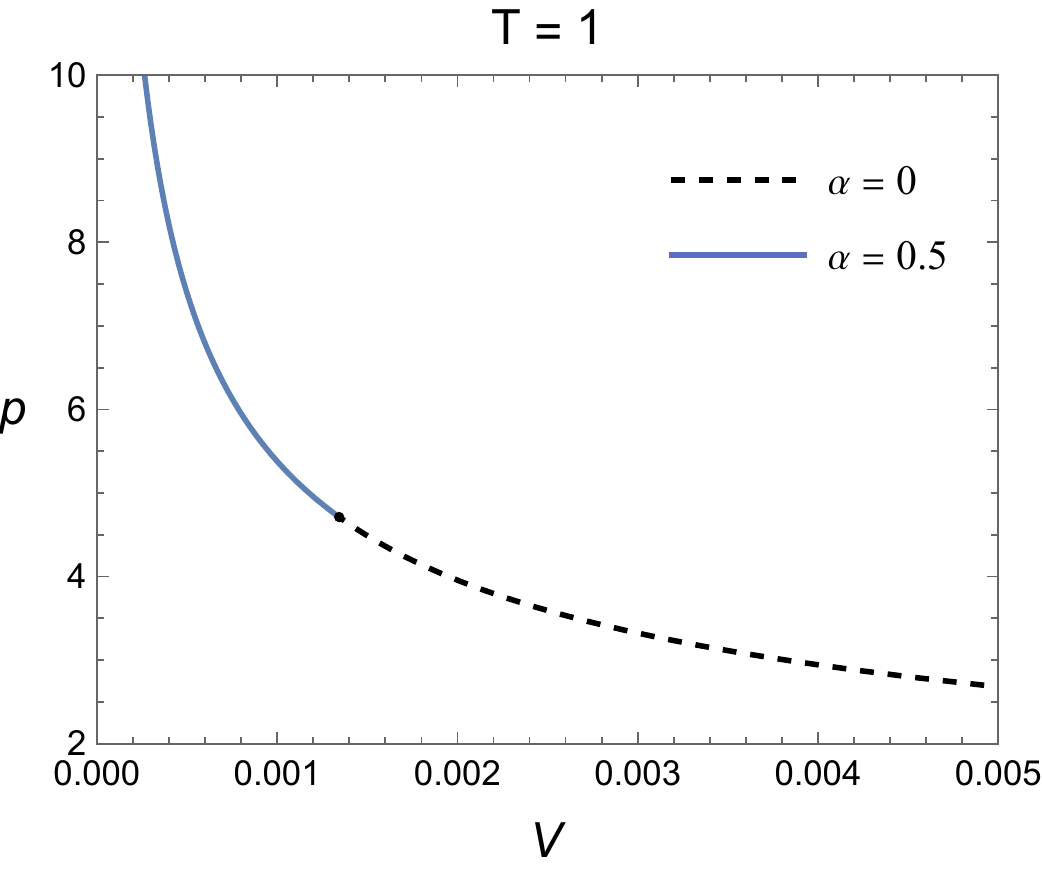}\includegraphics[width=7.6cm]{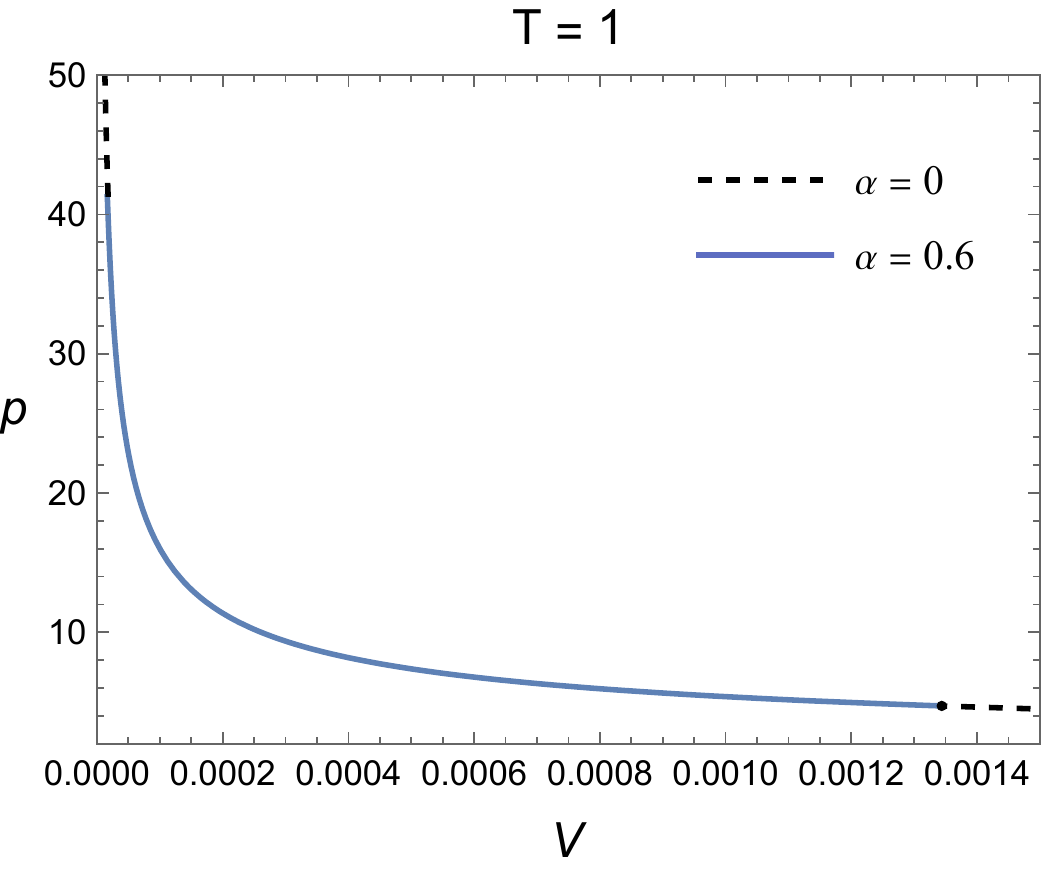}
\caption{
Isotherms in the $P$-$V$ diagrams of the hairy hyperbolic AdS black hole. We set $G_{N}=V_{\Sigma}=T=1$. Left: the solid curve is the hairy hyperbolic black hole with $\alpha=0.5$, while the dashed curve is the hyperbolic black hole with $\alpha=0$ (without scalar hair). Right: the solid curve is the hairy hyperbolic black hole with $\alpha=0.6$, while the dashed curve is the hyperbolic black hole with $\alpha=0$(without scalar hair).}
\label{fig:1}
\end{figure}
The $P$-$V$ diagram of the transition behaviors is given by Fig.~\ref{fig:1}. It shows that there is no van der Waals behaviors for the hairy hyperbolic AdS black hole.

\begin{figure}[htbp]
\centering
\includegraphics[width=7.6cm]{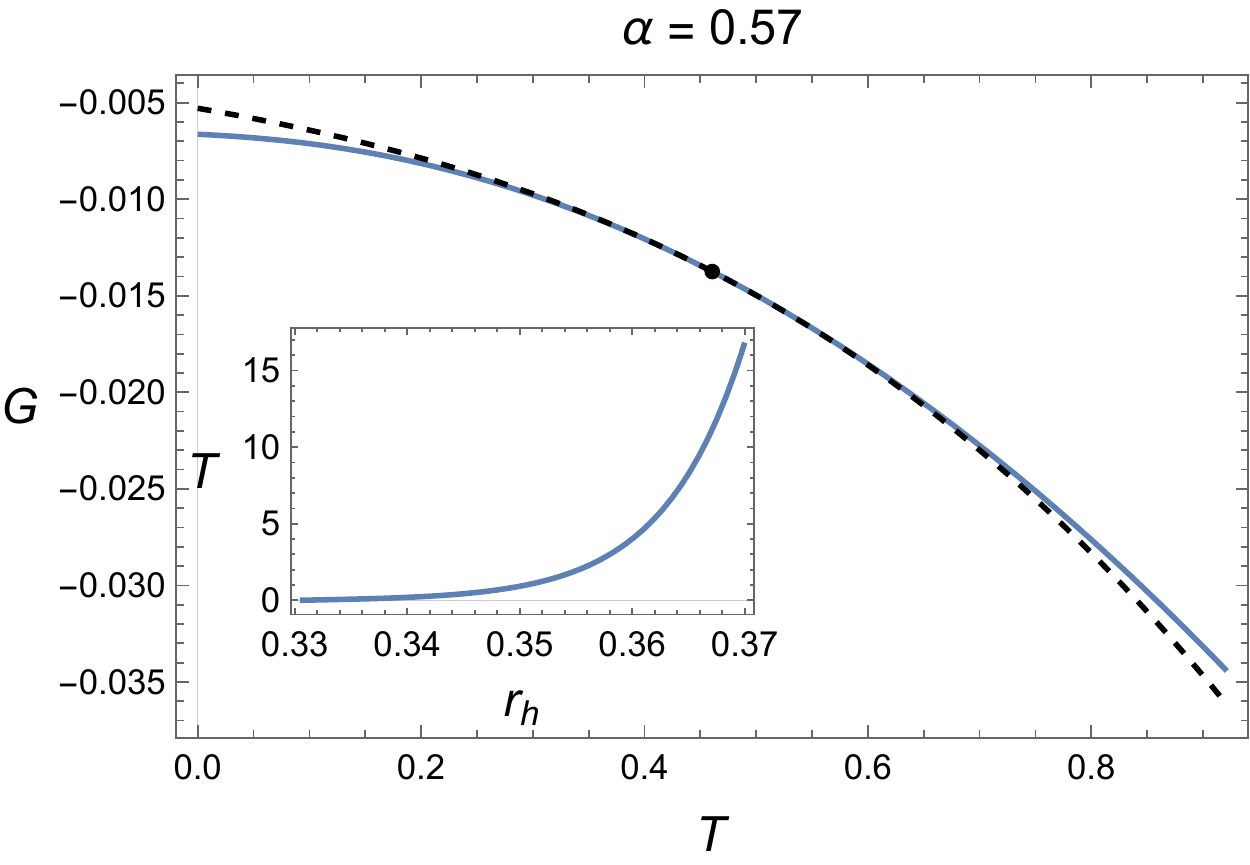}\includegraphics[width=7.6cm]{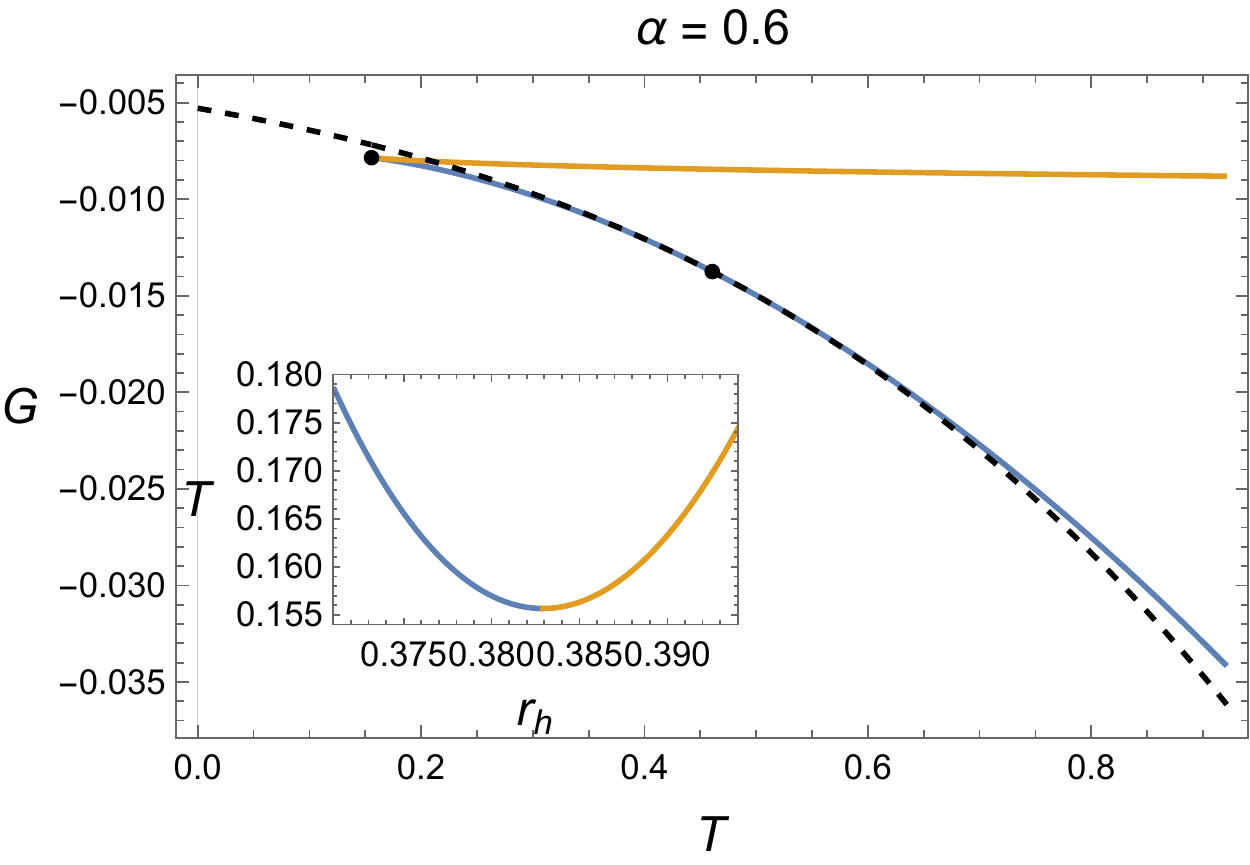}
\includegraphics[width=7.6cm]{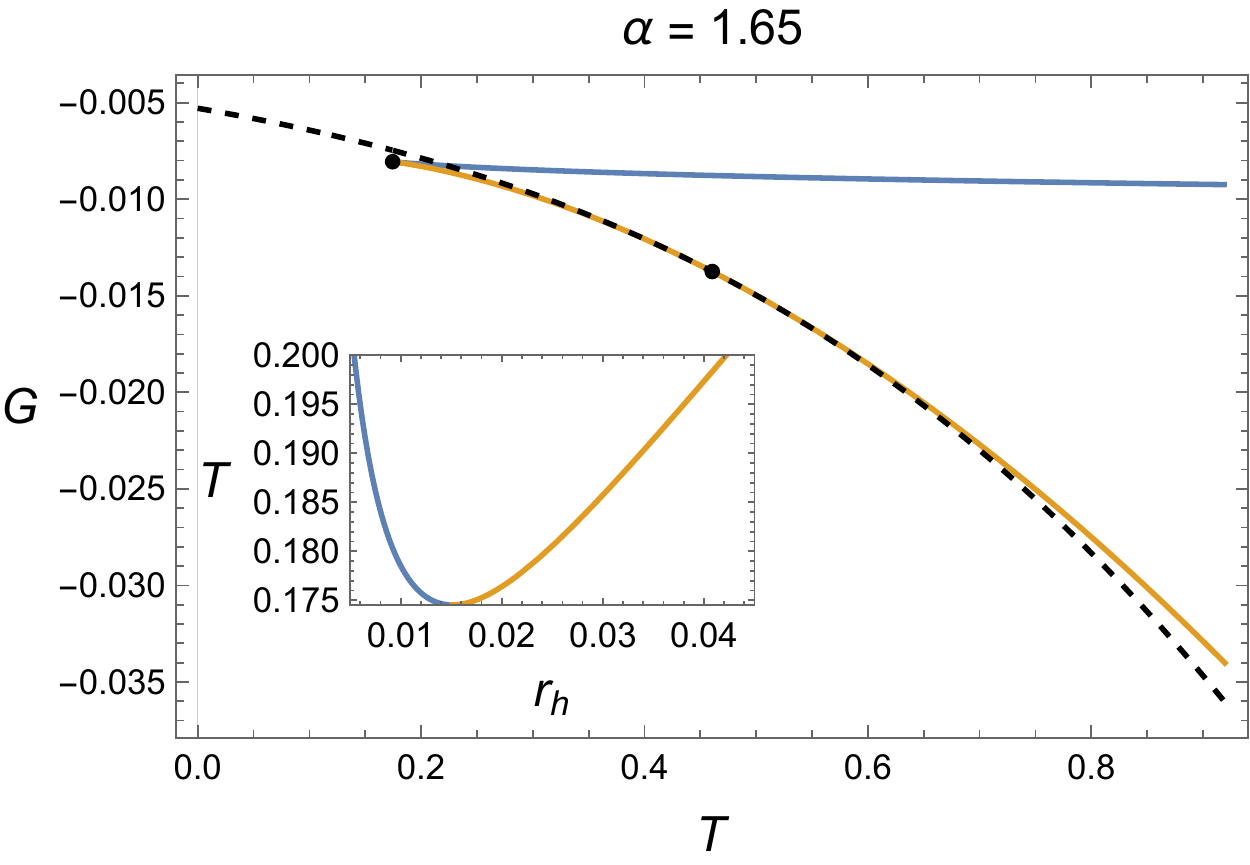}\includegraphics[width=7.6cm]{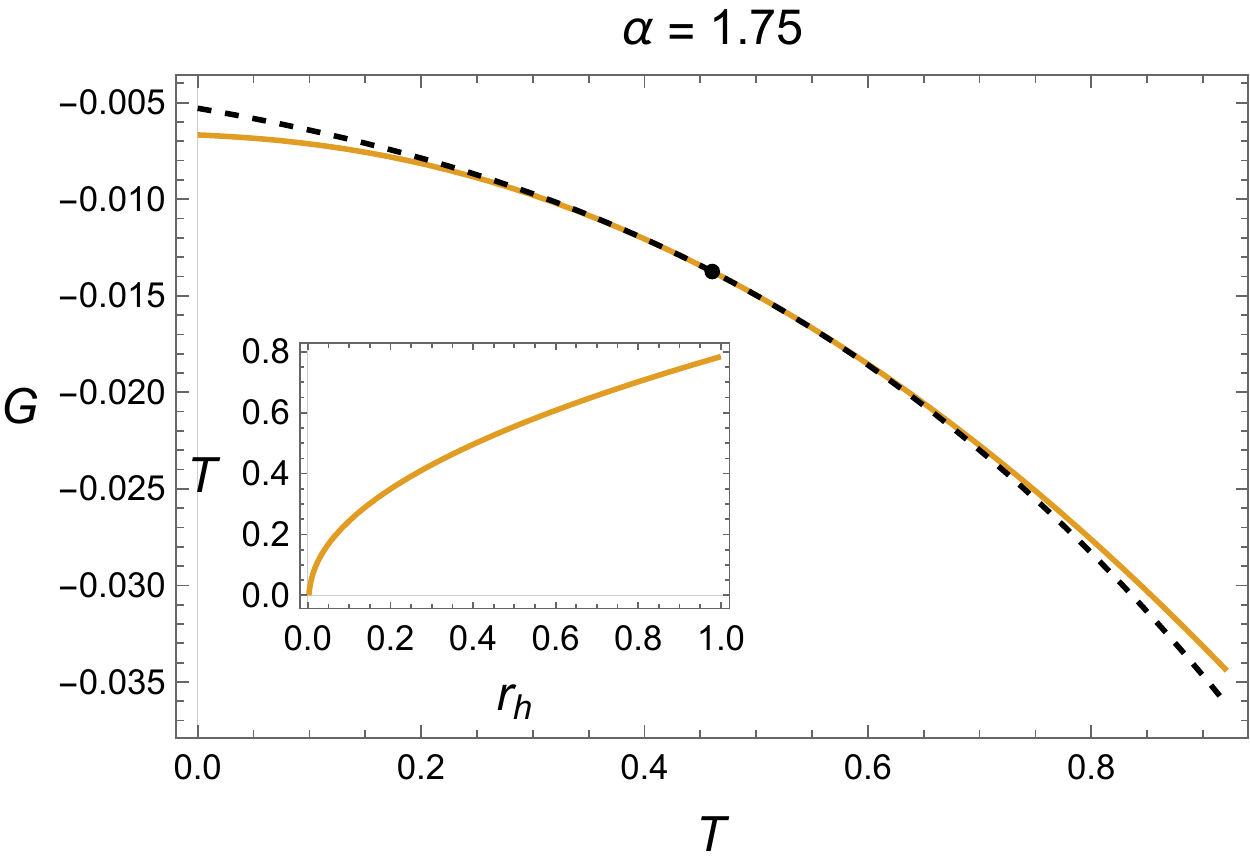}
\caption{
The Gibbs free energy as a function of temperature at fixed $L=G_{N}=V_{\Sigma}=1$ for different values of $\alpha$. The solid curve is the hairy black hole where blue one is for $r_{+}$, and orange one is for $r_{-}$, while the dashed curve is the hyperbolic black hole without scalar hair. The relation between the temperature and horizon radius of black hole is described in the subfigure.}
\label{fig:2}
\end{figure}

The Gibbs free energy is the Legendre transform of the enthalpy,
\begin{equation}
\label{eq:Gibbs-free}
  G=H-TS=-\frac{V_{\Sigma}}{16\pi G_{N}}r_{h}\left[1+\left(\frac{r_{h}}{L}\right)^{\frac{2(1+\alpha ^{2})}{1-3\alpha ^{2}}}\right].
\end{equation}
The Gibbs free energy is equal to the Helmholz free energy in original thermodynamics of the black hole. Therefore, we can directly use the thermodynamic quantities calculated in ~\cite{Bai:2022obp}. The behaviors of the Gibbs free energy is depicted Fig.~\ref{fig:2}.

\begin{figure}[htbp]
\centering
\includegraphics[width=7.6cm]{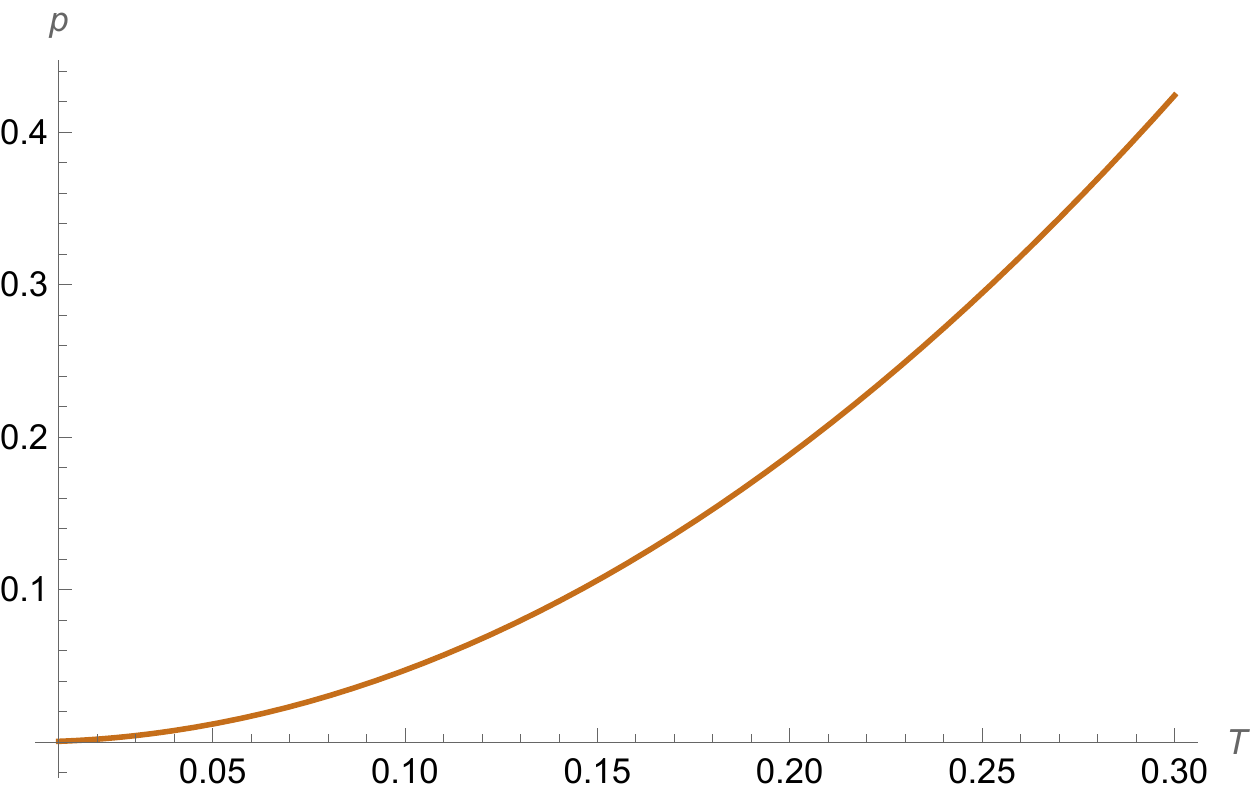}\includegraphics[width=7.6cm]{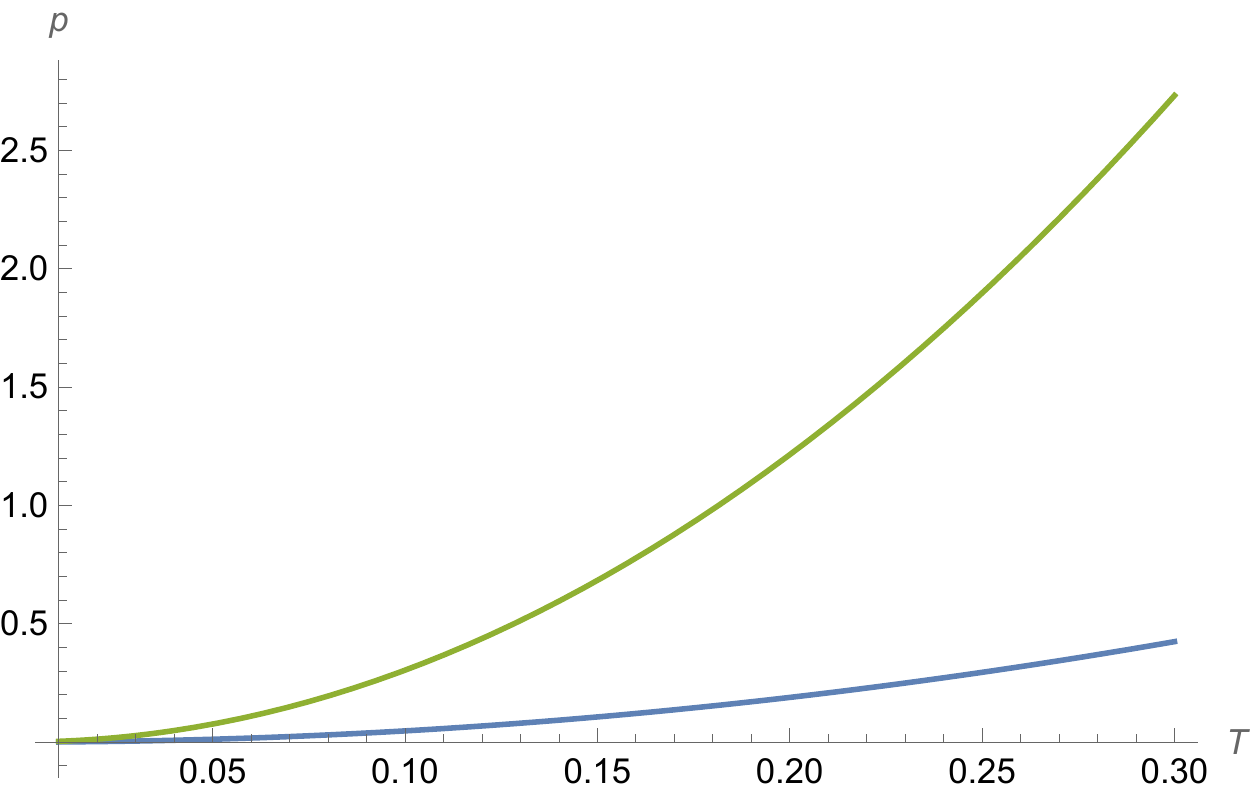}
\caption{
The phase diagram for the hairy hyperbolic AdS black hole. Left: the coexistence curve is for the hairy black hole with $\alpha=0.5$ and the black hole without scalar hair. Right: the coexistence curve is for the hairy black hole with $\alpha=0.6$ and the black hole without scalar hair. The heat capacity diverges on the lower(blue) line.}
\label{fig:3}
\end{figure}

For $0\le \alpha\le \frac{1}{\sqrt{3}}$, the hairy black hole has a horizon radius $r_{+}$ in eq.\eqref{eq:r+-}, and the third-order phase transition happens at critical temperature $T_{0}=1/(2\pi L_{0})$. The pressure $p\vert_{\text{coexistance}}$ can be computed as $p=\frac{3\pi}{2}T^{2}$ in terms of critical temperature $T_{0}$, the two states can exist together as shown in the $p$-$T$ plane in Fig.~\ref{fig:3}.

For $\alpha\ge \sqrt{3}$, the hairy black hole has a horizon radius $r_{-}$ in eq.\eqref{eq:r+-}, and the third-order phase transition happens at critical temperature $T_{0}=1/(2\pi L_{0})$, this $p$-$T$ diagram is similar to the case of $0\le \alpha\le \frac{1}{\sqrt{3}}$.

For $\frac{1}{\sqrt{3}}< \alpha < \sqrt{3}$, there are two solutions of black holes, i.e., both $r_{h}=r_{\pm}$ in eq.\eqref{eq:r+-} are real, and the hairy black hole has a minimum temperature
\begin{equation}
\label{eq:t-min}
T_{m}=\frac{\sqrt{(3-\alpha ^{2})(3\alpha^{2}-1)}}{2\pi(1+\alpha ^{2})L},
\end{equation}
the third-order phase transition happens at critical temperature $T_{0}=1/(2\pi L_{0})$, and the zeroth-order phase transition happens at critical temperature $T_{m}$, the corresponding to the pressure $p\vert_{\text{coexistance}}$ can be computed as  $p=\frac{3\pi(1+\alpha^{2})^{2}}{2(3-\alpha^{2})(3\alpha^{2}-1)}T^{2}$  in terms of critical temperature $T_{m}$. The two states can exist together as shown in the $p$-$T$ plane in Fig.~\ref{fig:3}.

The heat capacity can be expressed as~\cite{Bai:2022obp}
\begin{equation}
\label{eq:c-p}
C_{p}=T\frac{\partial S}{\partial T}\lvert_{p}=\pm \frac{V_{\Sigma}}{G_{N}}\frac{\pi(1-\alpha ^{2})L^{3}T}{\sqrt{4\pi^{2}L^{2}T^{2}(1+\alpha ^{2})^{2}+(3-\alpha ^{2})(1-3\alpha^{2})}}x_{q}^{\frac{2(1-\alpha ^{2})}{1-3\alpha ^{2}}}.
\end{equation}
Here $x_{q}=r_{h}/L$. From eq.\eqref{eq:c-p} and Fig.~\ref{fig:2}, we can see that the heat capacity diverges at $T_{m}$. For $\frac{1}{\sqrt{3}}<\alpha<1$, $r_{+}$ is on the lower branch, and the heat capacity $C_{p}>0$; while $r_{-}$ is on the upper branch, and $C_{p}<0$. Thus the more stable hairy black hole corresponds to $r_{+}$. For $1<\alpha<\sqrt{3}$, $r_{-}$ is on the lower branch, and the heat capacity $C_{p}>0$; while $r_{+}$ is on the upper branch, and $C_{p}<0$. Thus the more stable hairy black hole corresponds to $r_{-}$. When $\alpha=1$ the heat capacity $C_{p}=0$, and when $\alpha=0$ the heat capacity $C_{p}>0$, thus the black hole without scalar hair is more stable.

\section{Extended  R\'{e}nyi entropies}

In this section, we will give the extended holographic R\'{e}nyi entropy $S_{q,b}$ of the hairy black hole in the extended framework of the black hole thermodynamics. The usual holographic R\'{e}nyi entropy of the hairy black hole at fixed $L$, it can be written as ~\cite{Bai:2022obp}
\begin{equation}
\label{eq:sq-g}
S_{q}= \frac{q}{q-1}\frac{S_{EE}}{2}\left[2-x_{q}-x_{q}^{\frac{3-\alpha ^{2}}{1-3\alpha ^{2}}}\right],
\end{equation}
where
\begin{equation}
\label{eq:x-q}
x_{q}=r_{h}/L= \left(\frac{1+\alpha ^{2} \pm \sqrt{(1+\alpha ^{2})^{2}+(3-\alpha ^{2})(1-3\alpha ^{2})q^{2}}}{(3-\alpha ^{2})q} \right)^{\frac{1-3\alpha ^{2}}{1+\alpha ^{2}}},
\end{equation}
$S_{EE}=\frac{L_{0}^{2}}{4G_{N}}V_{\Sigma}$ corresponds to the entanglement entropy, and $r_{h}$ corresponds to the more stable black hole solutions: $r_{h}=r_{+}$ for $0 \le \alpha < 1$  , and $r_{h}=r_{-}$ for $\alpha > 1$.

In order to extend R\'{e}nyi entropy of the hairy black hole, we allow for pressure changes, let $L_{0}$ change to $L=\frac{L_{0}}{b}$.  Via eq.\eqref{eq:sqb1}, we obtain
\begin{equation}
\label{eq:sq-g}
S_{q,b}= \frac{q}{qb^{2}-1}\frac{S_{EE}}{2}\left[2-\frac{1}{b}(x_{q,b}+x_{q,b}^{\frac{3-\alpha ^{2}}{1-3\alpha ^{2}}})\right],
\end{equation}
where
\begin{equation}
\label{eq:x-q-b}
x_{q,b}= \left(\frac{1+\alpha ^{2} \pm \sqrt{(1+\alpha ^{2})^{2}+(3-\alpha ^{2})(1-3\alpha ^{2})q^{2}b^{2}}}{(3-\alpha ^{2})qb} \right)^{\frac{1-3\alpha ^{2}}{1+\alpha ^{2}}},
\end{equation}
when $\alpha=0$, $q\to 1$, $b\to 1$, we can see the extended R\'{e}nyi entropy $S_{q,b}$ reduces to the  entanglement entropy $S_{1,1}=S_{EE}$.

 \begin{figure}[htbp]
\centering
\includegraphics[width=7.8cm]{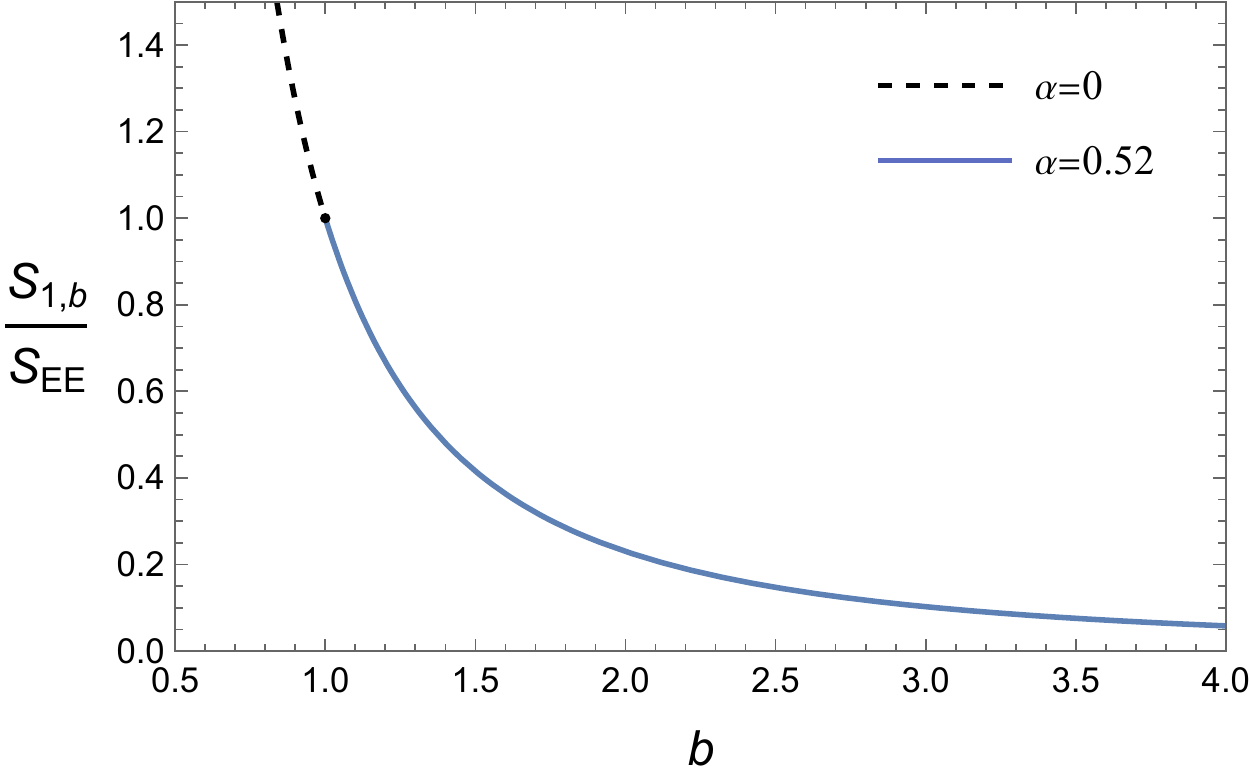}\includegraphics[width=7.8cm]{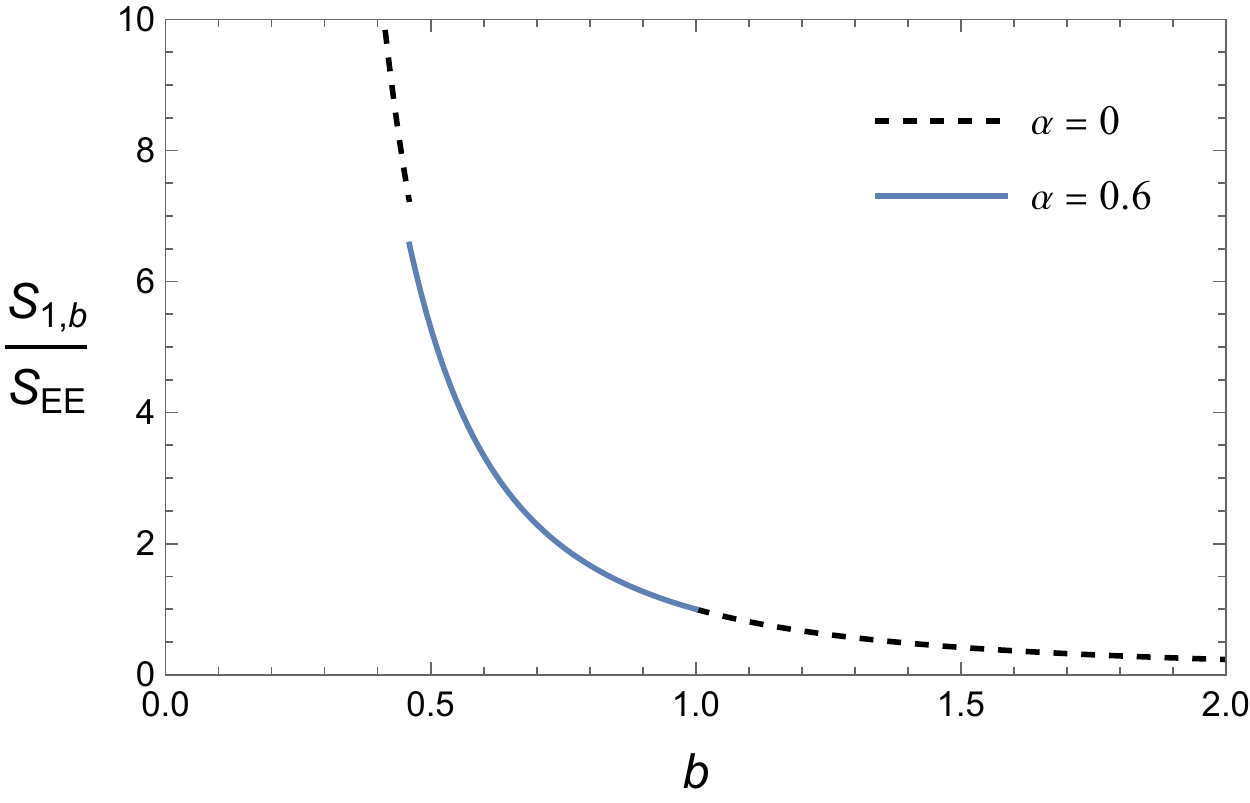}
\includegraphics[width=7.8cm]{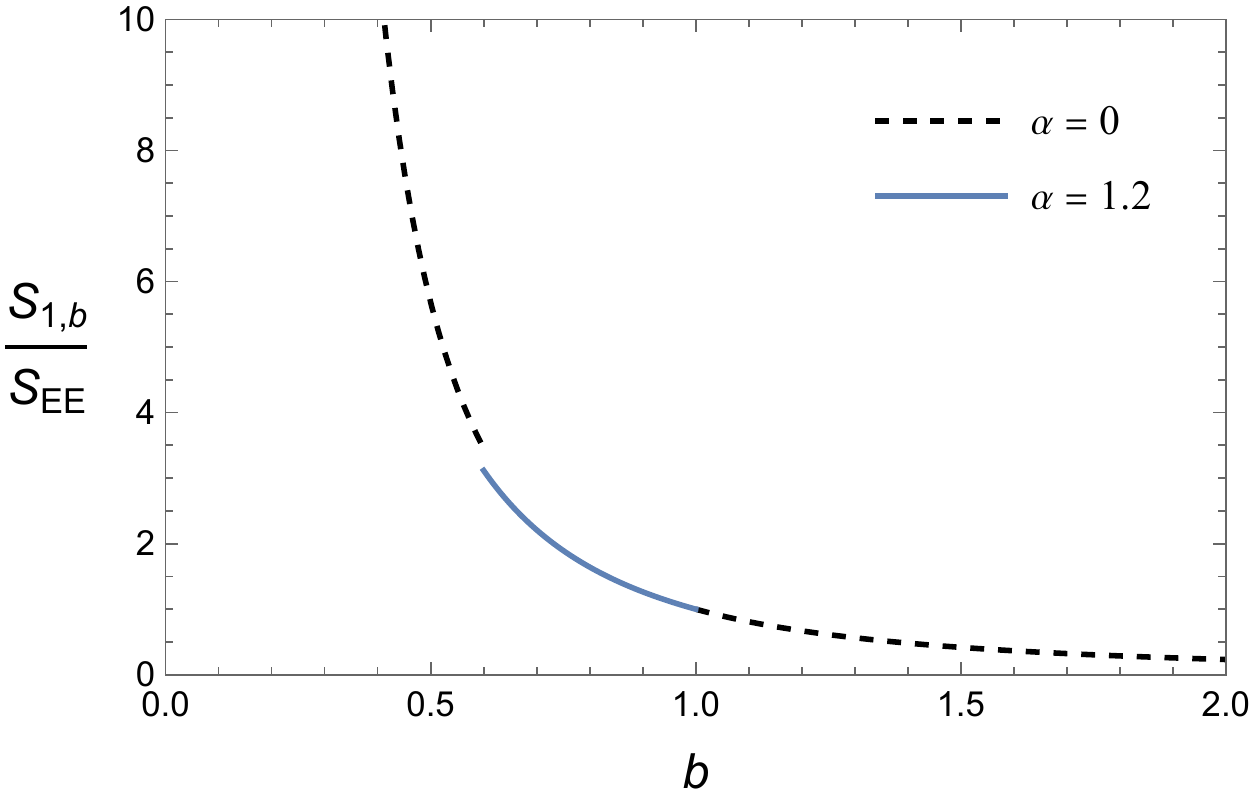}\includegraphics[width=7.8cm]{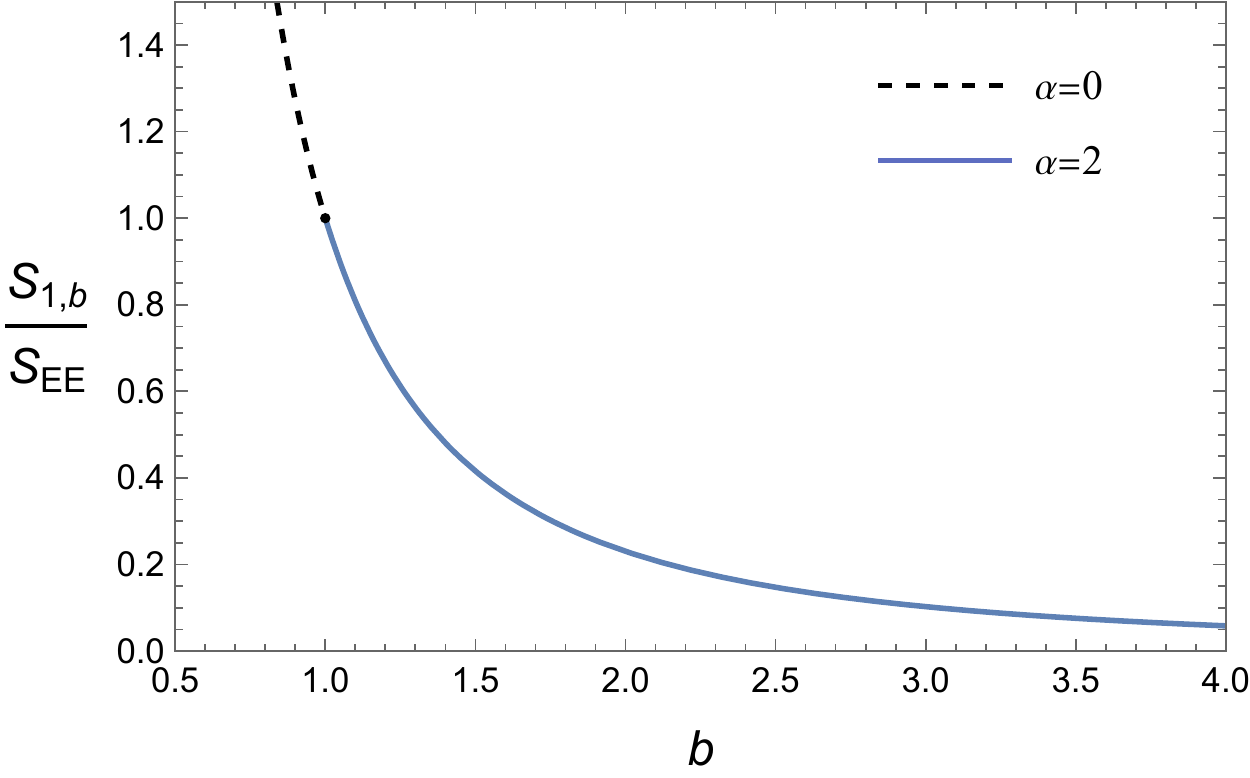}
\caption{
The extended R\'{e}nyi entropy $S_{1,b}$ as a function of $b$ at fix $q=1$. The solid curve is the hairy hyperbolic AdS black hole, while the dashed curve is the hyperbolic black hole without scalar hair.}
\label{fig:4}
\end{figure}

To show that the different behavior of the extended holographic R\'{e}nyi entropy at the critical temperature when the black hole has a thermodynamic phase transition. We can show that the extended R\'{e}nyi entropy $S_{1,b}$ as a function of $b$ at fix $q=1$ for different values of $\alpha$.

For $0 \le \alpha\le \frac{1}{\sqrt{3}}$, the hairy black hole corresponds to $r_{+}$, the third-order phase transition happens at critical point $q=1, b=1$. The relationship between  $S_{1,b}$  and $b$ is shown in top left of Fig.~\ref{fig:4}.

For $\alpha\ge \sqrt{3}$, the hairy black hole corresponds to $r_{-}$, the third-order phase transition happens at critical point $q=1, b=1$. The relationship between  $S_{1,b}$  and $b$ is shown in bottom right of Fig.~\ref{fig:4}.

When $\frac{1}{\sqrt{3}}<\alpha< 1$, the third-order phase transition happens at critical point $q=1, b=1$, and the zeroth-order phase transition happens at critical point
\begin{equation}
\label{eq:b-m}
 b_{m}=\frac{1+\alpha ^{2}}{q_{m}\sqrt{(3-\alpha ^{2})(3\alpha^{2}-1)}},
 \end{equation}
with $q_{m}=T_{0}/T_{m}$, the relationship between  $S_{1,b}$ and $b$ is shown in the top right of Fig.~\ref{fig:4}. We can see that when the zeroth-order phase transition of the black hole happens it leads to the extended holographic R\'{e}nyi entropy to a finite jump at the critical point $b_{m}$. For $1<\alpha< \sqrt{3}$, it is similar to the case of $\frac{1}{\sqrt{3}}<\alpha< 1$ and the relationship between the $S_{1,b}$ and $b$ is shown in the bottom left of Fig.~\ref{fig:4}.

The R\'{e}nyi entropy has been used to explore information theory and condensed matter physics~\cite{PhysRevLett.101.010504,Flammia:2009axf,Metlitski:2009iyg,Headrick:2010zt}.  Recall that the limits of the R\'{e}nyi entropy relate to the entanglement spectrum. In the $q \to \infty$ limit of eq.\eqref{eq:renyi}, we have  $S_{\infty}=\displaystyle\lim_{q \to \infty }S_{q}=-\text{log}\lambda_{1}$, where $\lambda_{1}$ is the largest eigenvalue of $\rho_{A}$, and as $q \to 0$ we have $S_{0}=\displaystyle\lim_{q \to 0}S_{q}=\text{log} (d)$, where $d$ corresponds to the number of nonvanishing eigenvalues of $\rho_{A}$.

Now, let us consider some important limits for extended R\'{e}nyi entropy $S_{q,b}$ from the hairy black hole. First, we note that when $0\leqslant \alpha \leqslant \frac{\sqrt{3}}{3}$ or $\alpha \geqslant \sqrt{3}$
\begin{equation}
\label{eq:x-m-inf}
\displaystyle\lim_{q \  \text{or} \  b \to \infty}x_{q,b} \to \left(\frac{1-3\alpha^{2}}{3-\alpha^{2}}\right)^{\frac{1-3\alpha^{2}}{2(1+\alpha^{2})}},
 \end{equation}
while for $\frac{\sqrt{3}}{3}<\alpha<\sqrt{3}$, as $q \  \text{or} \  b \to \infty$ the $x_{q,b}$ is not real.

For $0\leqslant\alpha<1$, as  $q \  \text{or} \  b \to 0$ we have
\begin{equation}
\label{eq:x-0-a1}
\displaystyle\lim_{q \  \text{or} \  b \to 0}x_{q,b} \to \left[\frac{2(1+\alpha^{2})}{(3-\alpha^{2})qb}\right]^{\frac{1-3\alpha^{2}}{1+\alpha^{2}}},
 \end{equation}
for $\alpha>1$, we have
\begin{equation}
\label{eq:x-a-3}
\displaystyle\lim_{q \  \text{or} \  b \to 0}x_{q,b} \to \left[\frac{2(1+\alpha^{2})}{(3\alpha^{2}-1)qb}\right]^{\frac{3\alpha^{2}-1}{1+\alpha^{2}}}.
 \end{equation}

For $0\leqslant \alpha \leqslant \frac{\sqrt{3}}{3}$ or $\alpha \geqslant \sqrt{3}$ at the fix $b$, by taking the $q \to \infty$ limit of extended R\'{e}nyi entropy $S_{q,b}$ we have
\begin{equation}
\label{eq:s-q-infty}
\displaystyle\lim_{q \to \infty}S_{q,b} = \frac{S_{EE}}{2b^{2}}\left[2-\frac{1}{b}\left(\frac{1-3\alpha^{2}}{3-\alpha^{2}}\right)^{\frac{1-3\alpha^{2}}{2(1+\alpha^{2})}}+\left(\frac{1-3\alpha^{2}}{3-\alpha^{2}}\right)^{\frac{3-\alpha^{2}}{2(1+\alpha^{2})}}\right],
 \end{equation}
and at the fix $q$, and the limit $b \to \infty$ we have
\begin{equation}
\label{eq:s-b-infty}
\displaystyle\lim_{b \to \infty}S_{q,b} = \frac{S_{EE}}{b^{2}}.
 \end{equation}

For $0\leqslant\alpha<1$ and as $q \to 0$ at the fix $b$, we have
\begin{equation}
\label{eq:s-q-03}
\displaystyle\lim_{q \to 0}S_{q,b} =\frac{S_{EE}}{2b}\left[\frac{2(1+\alpha^{2})}{(3-\alpha^{2})b}\right]^{\frac{3-\alpha^{2}}{1+\alpha^{2}}}q^{\frac{2(\alpha^{2}-1)}{1+\alpha^{2}}}.
\end{equation}

For $0\leqslant\alpha<1$ and as $b \to 0$ at the fix $q$, we have
\begin{equation}
\label{eq:s-b-03}
\displaystyle\lim_{b \to 0}S_{q,b} =\frac{qS_{EE}}{2}\left[\left(\frac{2(1+\alpha^{2})}{(3-\alpha^{2})q}\right)^{\frac{1-3\alpha^{2}}{1+\alpha^{2}}}b^{\frac{2(\alpha^{2}-1)}{1+\alpha^{2}}}+\left(\frac{2(1+\alpha^{2})}{(3-\alpha^{2})q}\right)^{\frac{3-\alpha^{2}}{1+\alpha^{2}}}b^{-\frac{4}{1+\alpha^{2}}}\right].
\end{equation}

For $\alpha >1$ and as $q \to 0$ at the fix $b$, we have
\begin{equation}
\label{eq:s-q-3-3}
\displaystyle\lim_{q \to 0}S_{q,b} =\frac{S_{EE}}{2b} \left[\frac{2(1+\alpha^{2})}{(3\alpha^{2}-1)b}\right]^{\frac{3\alpha^{2}-1}{1+\alpha^{2}}}q^{\frac{2(1-\alpha^{2})}{1+\alpha^{2}}}.
\end{equation}

For $\alpha >1$ and as $b \to 0$ at the fix $q$, we have
\begin{equation}
\label{eq:s-b-3-3}
\displaystyle\lim_{b \to 0}S_{q,b} =\frac{qS_{EE}}{2}\left[\left(\frac{2(1+\alpha^{2})}{(3\alpha^{2}-1)q}\right)^{\frac{3\alpha^{2}-1}{1+\alpha^{2}}}b^{-\frac{4\alpha^{2}}{1+\alpha^{2}}} +\left(\frac{2(1+\alpha^{2})}{(3\alpha^{2}-1)q}\right)^{\frac{\alpha^{2}-3}{1+\alpha^{2}}}b^{\frac{2(1-\alpha^{2})}{1+\alpha^{2}}}\right].
\end{equation}

The usual  R\'{e}nyi entropy is known to satisfy following inequalities~\cite{Hung:2011nu}
 \begin{equation}
  \begin{aligned}
  \label{eq:ineq}
 \frac{\partial S_{q}}{\partial q}& \le 0, \  \  \     \frac{\partial}{\partial q}\left(\frac{q-1}{q} S_{q}\right)&\ge 0, \\
 \frac{\partial}{\partial q}[(q-1) S_{q}]&\ge 0, \ \ \    \frac{\partial^{2}}{\partial q^{2}}[(q-1) S_{q}]&\le 0.
  \end{aligned}
  \end{equation}
 We will show that $S_{1,b}$ satisfies similar inequalities in our holographic calculations. In the following, we consider three such inequalities:
\begin{equation}
  \label{eq:eqs2}
 \frac{\partial}{\partial b}[(b^{2}-1) S_{1,b}] \ge 0.
 \end{equation}
\begin{equation}
  \label{eq:eqs1}
 \frac{\partial S_{1,b}}{\partial b} \le 0.
\end{equation}
 \begin{equation}
  \label{eq:eqs3}
 \frac{\partial^{2}}{\partial b^{2}}[(b^{2}-1) S_{1,b}] \le 0.
  \end{equation}
First, let us recall the eq.\eqref{eq:sqb1} which relates  the extended R\'{e}nyi entropy and thermodynamic quantities. 
When $q=1$ and $b \ne 1$ we know that $\Delta T=0$, $\Delta p \ne 0$, and then the eq.\eqref{eq:sqb1} can be written as
\begin{equation}
\label{eq:s1b-2}
S_{1,b}=\frac{S_{0}}{V_{0}\Delta p}\int_{p_{0}}^{b^{2}p_{0}}V(p)dp=\frac{S_{0}}{V_{0}p_{0}(b^{2}-1)}\int_{p_{0}}^{b^{2}p_{0}}V(p)dp.
\end{equation}

If we begin by considering the inequality \eqref{eq:eqs2}, the expression on the right-hand side yields
\begin{equation}
\frac{\partial}{\partial b}[(b^{2}-1) S_{1,b}]=\frac{2S_{0}b}{V_{0}}\frac{\partial}{\partial p}\int_{p_{0}}^{b^{2}p_{0}}V(p)dp=\frac{2S_{0}bV(b^{2}p_{0})}{V_{0}}\ge 0.
\end{equation}

Next turning to the inequality \eqref{eq:eqs1}, from eq.\eqref{eq:s1b-2}, we can get
\begin{equation}
\label{eq:into-2}
\frac{\partial S_{1,b}}{\partial b}=\frac{2S_{0}b}{V_{0}p_{0}(b^{2}-1)^{2}} \int_{p_{0}}^{b^{2}p_{0}}\left[V(b^{2}p_{0})-V(p)\right]dp.
\end{equation}
From the stability requirement, the integrand above is negative for $b > 1$.  For $0 < b < 1$, the above integrand eq.\eqref{eq:into-2} is rewritten as
\begin{equation}
\label{eq:into-3}
\frac{\partial S_{1,b}}{\partial b}=\frac{2S_{0}b}{V_{0}p_{0}(b^{2}-1)^{2}} \int_{b^{2}p_{0}}^{p_{0}}\left[V(p)-V(b^{2}p_{0})\right]dp.
\end{equation}
The same reasoning of the above integrand eq.\eqref{eq:into-3} is still satisfied for this range of $b$. Thus, the inequality \eqref{eq:eqs1} is satisfied.

Finally, let us consider the inequality \eqref{eq:eqs3}, the right-hand side yields
\begin{equation}
\label{eq:ineq3}
\frac{\partial^{2}}{\partial b^{2}}[(b^{2}-1) S_{1,b}]=\frac{2S_{0}}{V_{0}}\left[V(b^{2}p_{0})+2b^{2}p_{0}\left(\frac{\partial V}{\partial p}\right)_{T}\right]=-\frac{2S_{0}V(b^{2}P_{0})}{V_{0}}(2b^{2}P_{0}\kappa_{T}-1).
\end{equation}
From eq.\eqref{eq:ineq3}, we can see that if the inequality \eqref{eq:eqs3} is satisfied, then the isothermal compressibility coefficient $\kappa_{T}$ need satisfy $\kappa_{T}=-\frac{1}{V}\left(\frac{\partial V}{\partial p}\right)_{T} \ge \frac{1}{2b^{2}P_{0}}$. 
For the hyperbolic black hole with scalar hair, we can verify that it is satisfied.

\section{Holographic computations for twist operators}

The calculation of the R\'{e}nyi entropy can be achieved by inserting a twist operator $\sigma_{q}$ at the entangling surface~\cite{Calabrese:2009qy,Calabrese:2004eu,Klebanov:2011uf,Calabrese:2005zw,Hung:2011nu}. The calculation of $\text{Tr} \rho_{A}^{q}$  is equivalent to the correlation function of the twist operators, thus the problem boils down to computing the conformal dimension of the twist operators.
In the paper~\cite{Hung:2011nu}, it shows the holographic computation of the conformal dimension $h_{q}$ of the twist operators in terms of gravitational quantities. Generically, the conformal dimension $h_{q}$ can be given in terms of the energy density $\varepsilon(T)$ of the boundary field theory. The conformal dimension $h_{q}$ of the twist operators in the four dimensions can be written as
\begin{equation}
\label{eq:gsh-q0}
h_{q}=\pi qL_{0}^{3}[\varepsilon(T_{0})-\varepsilon(T_{0}/q)]=\frac{\pi qL_{0}}{V_{\Sigma}}[M(T_{0})-M(T_{0}/q)],
\end{equation}
where $M$ is the dual black hole mass. For the twist operators $\sigma_{q}$, from eq.\eqref{eq:thermodynamics-quant}, we can obtain the conformal dimension
\begin{equation}
\label{eq:h-q-1}
h_{q}=\frac{qL_{0}^{2}}{8G_{N}}\frac{1-\alpha ^{2}}{1+\alpha ^{2}}\left[x_{q}-x_{q}^{\frac{3-\alpha ^{2}}{1-3\alpha ^{2}}}\right],
\end{equation}
where $x_{q}$ is given by eq.\eqref{eq:x-q}. Instead of staying at constant pressure, we can use the extended framework of the hairy black hole thermodynamics. For the twist operators $\sigma_{q}^{(b)}$, the conformal dimension can be obtained as
\begin{equation}
\label{eq:h-q-b-2}
h_{q}^{(b)}= \frac{qL_{0}^{2}}{8bG_{N}}\frac{1-\alpha ^{2}}{1+\alpha ^{2}}\left[x_{q,b}-x_{q,b}^{\frac{3-\alpha ^{2}}{1-3\alpha ^{2}}}\right],
\end{equation}
where $x_{q,b}$ is given by eq.\eqref{eq:x-q-b}. We can see that the conformal dimension $h_{q}^{(b)}$ can be given as a difference of enthalpy in the extended thermodynamics of black holes.

\section{Holographic extended capacity of entanglement}
The capacity of entanglement~\cite{PhysRevLett.105.080501} as an important quantum information measure was originally introduced to characterize topologically ordered states in the condensed matter physics and a fraction of works has been done in holography~\cite{Nakaguchi:2016zqi}.
In the earlier work~\cite{PhysRevLett.105.080501}, the capacity of entanglement was defined as
\begin{equation}
\label{eq:c-define}
C_{E}(q)=-t\frac{\partial^{2}}{\partial t^{2}}[(1-t)S_{1/t}],
\end{equation}
where $t=1/q$. For $q \to 1$, $C_{E}(1)=\text{Tr}\left[\rho_{A}(-\text{log}\rho_{A})^{2}\right]-\left[-\text{Tr}(\rho_{A}\text{log}\rho_{A})\right]^{2}$ gives the quantum fluctuation with respect to the original state $\rho_{A}$, the eq.\eqref{eq:c-define} can be rewritten as
\begin{equation}
\label{eq:c-deform}
C_{E}(1)=\displaystyle\lim_{q\to 1}q^{2}\partial_{q}^{2}\text{logTr}\rho_{A}^{q}.
\end{equation}
It is natural to generalize the capacity of entanglement in terms of the extended R\'{e}nyi entropy. Now, we define the extended capacity of entanglement as
\begin{equation}
\label{eq:c-qb1}
C_{E}(q,b)=-T\frac{\partial^{2}}{\partial T^{2}}\left[T(qb^{2}-1)S_{q,b}\right],
\end{equation}
where $T=1/q$. After a little algebra, the extended capacity of entanglement can be rewritten as
\begin{equation}
\label{eq:c-def-qb2}
C_{E}(q,b)= q^{2}\partial_{q}^{2}\text{log}[\text{Tr}(\rho_{A}^{(b)})^{q}],
\end{equation}
where
\begin{equation}
\label{eq:c-log}
\text{log}[\text{Tr}(\rho_{A}^{(b)})^{q}]=(1-qb^{2})S_{q,b}.
\end{equation}
We can see that when the $b=1$ the extended capacity of entanglement reduces to the usual capacity of entanglement.

Now, we show that the extended capacity of entanglement $C_{E}(q,b)$ of ball-like the subsystem $A$ maps to heat capacity $C_{\text{therm}}$ of the thermal CFT. From eqs.\eqref{eq:c-def-qb2} and \eqref{eq:c-log} without taking limits $ q\to 1, b\to 1$, we denote
\begin{equation}
\label{eq:cqb}
C_{E}(q,b)=q^{2}\partial_{q}^{2}\left[(1-qb^{2})S_{q,b}\right].
\end{equation}
The extended R\'{e}nyi entropy $S_{q,b}$ is related to the Gibbs free energy $G$ in the extended thermodynamics as
\begin{equation}
\label{eq:sqb--g}
S_{q,b}=-\frac{q}{T_{0}}\frac{[G(p_{0},T_{0})-G(b^{2}p_{0},T_{0}/q)]}{qb^{2}-1}.
\end{equation}

So
\begin{equation}
\label{eq:cqb--g}
C_{E}(q,b)=q^{2}\partial_{q}^{2}\left[-\frac{q}{T_{0}}G(b^{2}p_{0},T_{0}/q)\right],
\end{equation}
we can get
\begin{equation}
\label{eq:cqb--g}
C_{E}(q,b)=-\beta^{2}\frac{\partial H(T,b^{2}p_{0})}{\partial \beta}=\frac{\partial H(T,b^{2}p_{0})}{\partial T}=C_{\text{therm}},
\end{equation}
where $\beta=1/T$, the temperature $T=T_{0}/q$, and $H(T,b^{2}p_{0})$ is enthalpy. Thus when $q\to 1, b\to 1$, 
\begin{equation}
\label{eq:c-qb-th}
C_{E}(1,1)=\displaystyle\lim_{q\to 1, b\to 1}C(q,b)=\frac{\partial H(T_{0},p_{0})}{\partial T_{0}}=C_{\text{therm}}.
\end{equation}
For $\rho_{A}^{(b)}$, the extended capacity of entanglement $C_{E}(q,b)$ becomes heat capacity $C_{\text{therm}}$ of the thermal CFT on hyperbolic space. Its holographic dual interpretation is the heat capacity of the event horizon of the topological black hole in the bulk.

 \begin{figure}[htbp]
\centering
\includegraphics[width=7.6cm]{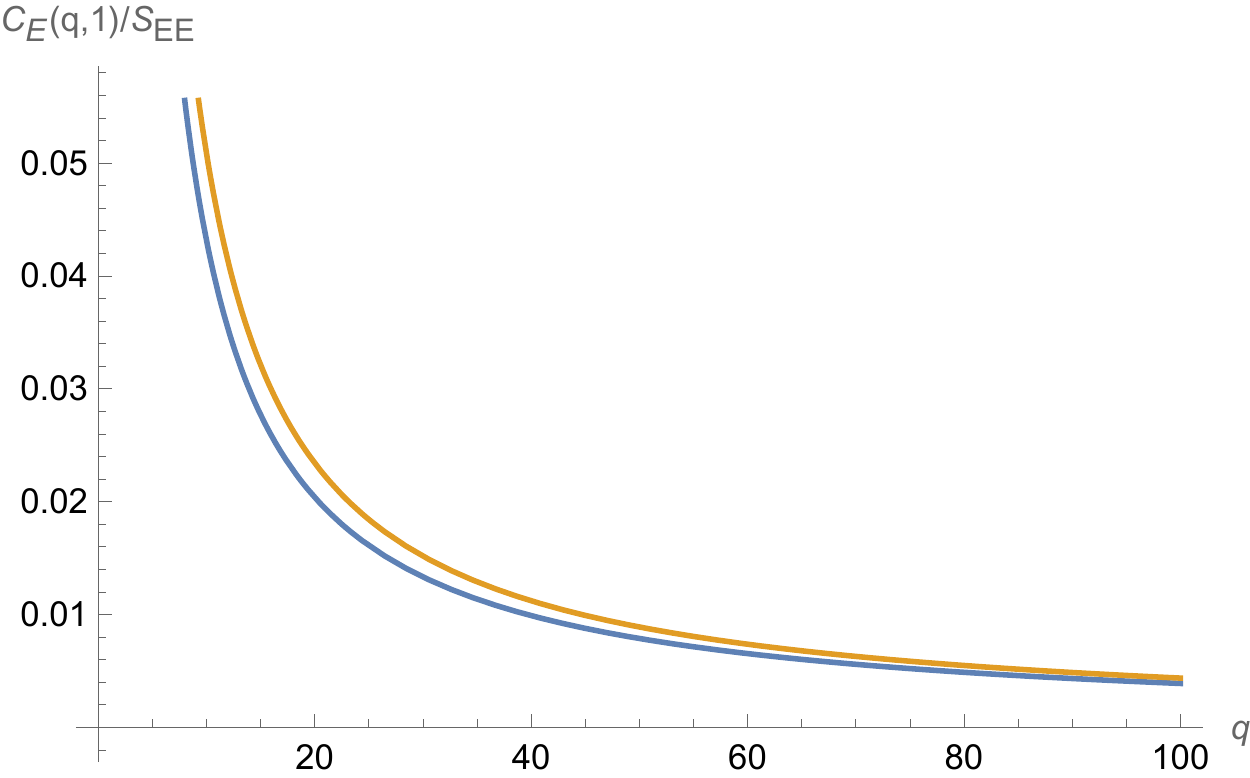}\includegraphics[width=7.6cm]{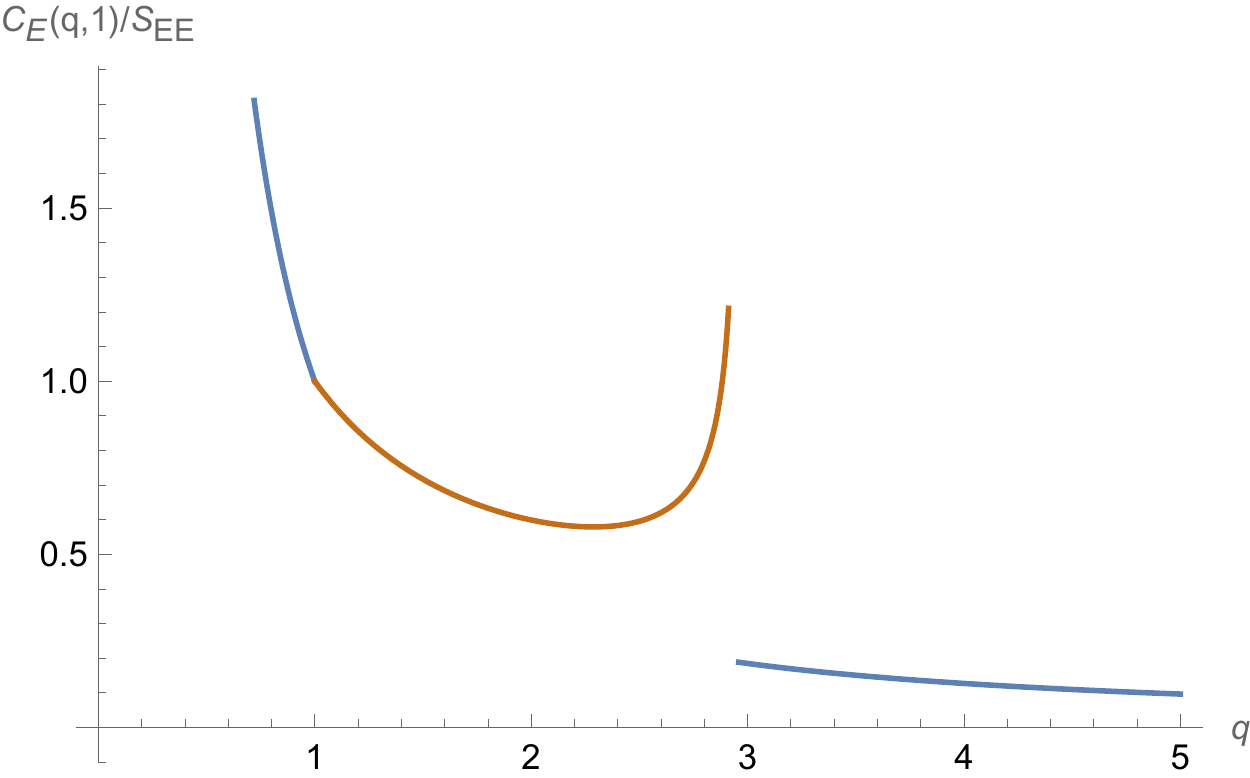}
\includegraphics[width=7.6cm]{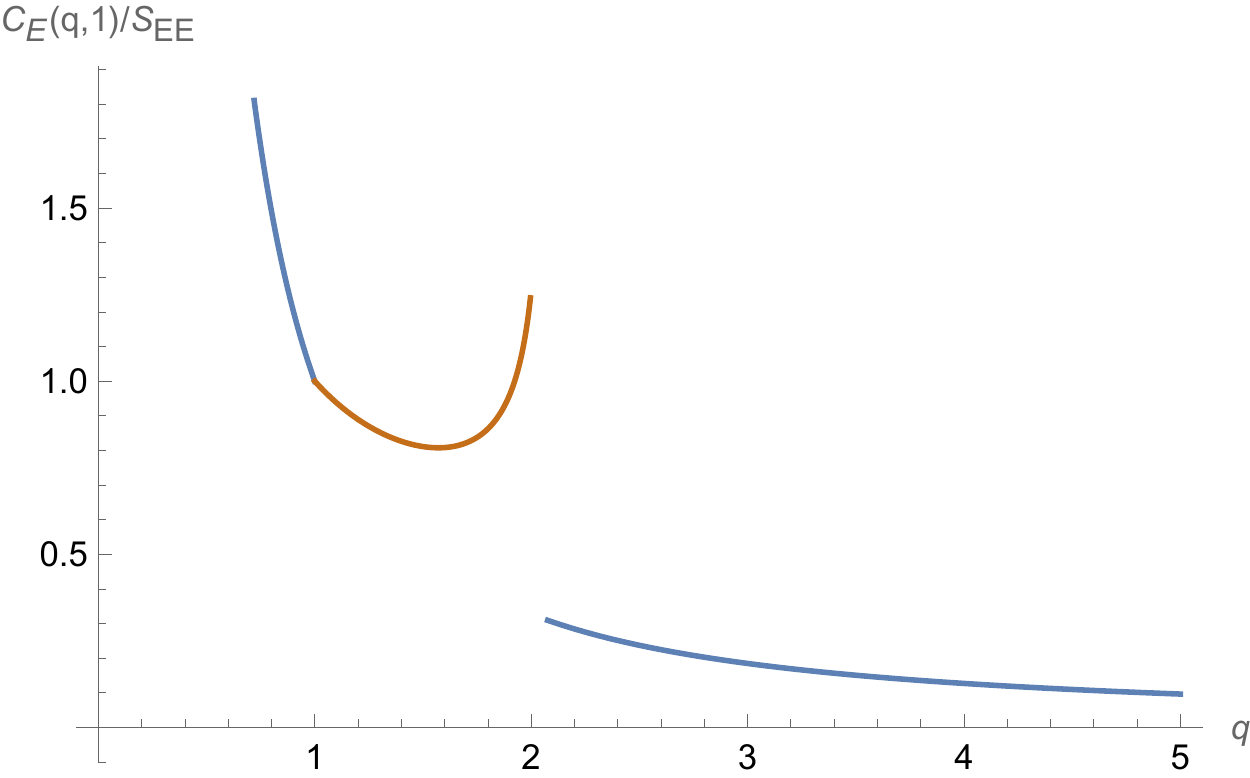}\includegraphics[width=7.6cm]{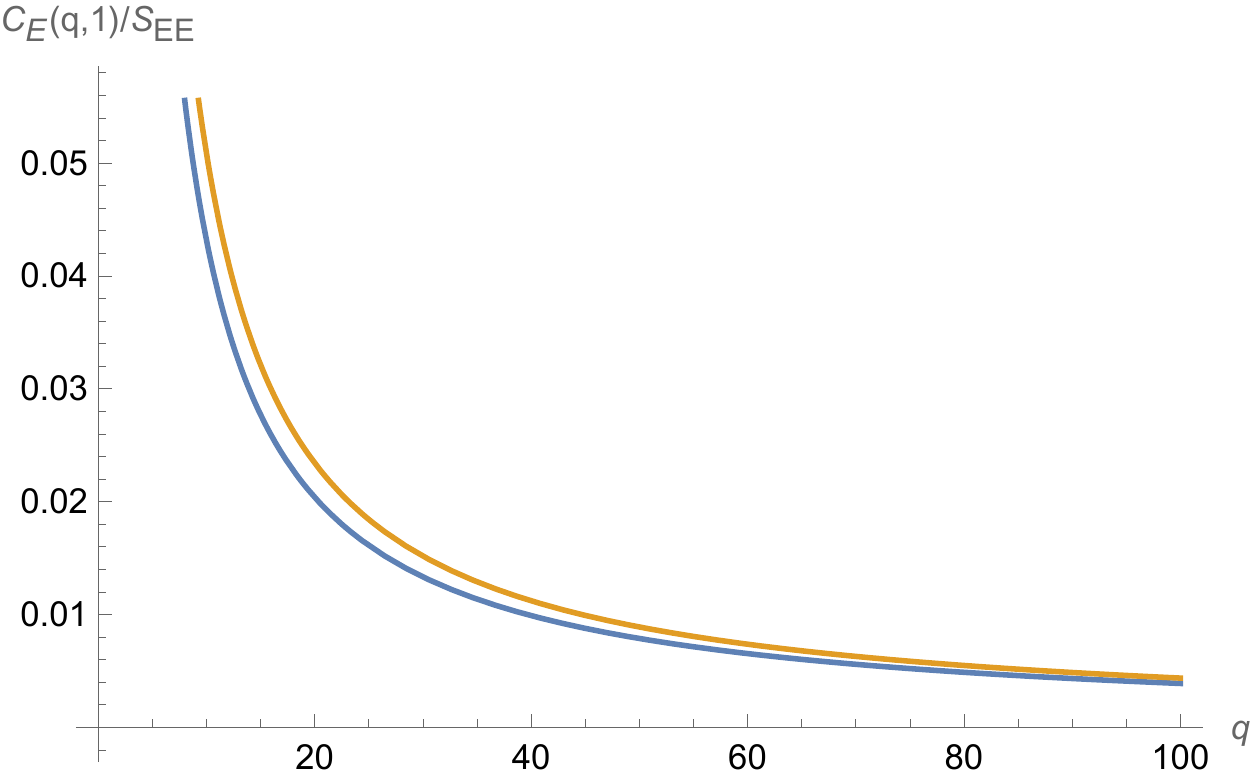}
\caption{
The extended capacity of entanglement $C_{E}(q,1)$ as a function of $q$ at fix $b=1$. The orange curve is the hairy hyperbolic AdS black hole, while the blue curve is the hyperbolic black hole without scalar hair.}
\label{fig:5}
\end{figure}
As an example of the extended capacity of entanglement, the eq.\eqref{eq:c-p} can be rewritten as
\begin{footnotesize}
\begin{equation}
\label{eq:c-qb}
C_{E}(q,b)=\pm \frac{\frac{2}{b^{3}q}(1-\alpha ^{2})S_{EE}}{\sqrt{\frac{(1+\alpha ^{2})^{2}}{b^{2}q^{2}}+(3-\alpha ^{2})(1-3\alpha^{2})}}\left(\frac{1+\alpha ^{2}\pm\sqrt{(1+\alpha ^{2})^{2}+(3-\alpha ^{2})(1-3\alpha^{2})b^{2}q^{2}}}{(3-\alpha ^{2})qb}\right)^{\frac{2(1-\alpha ^{2})}{1+\alpha ^{2}}}.
\end{equation}
\end{footnotesize}
For $\alpha=0$, we can see that $C_{E}(1,1)=S_{EE}$. The results for the extended capacity of entanglement \eqref{eq:c-qb} are illustrated in Fig.~\ref{fig:5}, which plot $C_{E}(q,1)/S_{EE}$ as a function of $q$ for various values of $\alpha$. Because the figure of $C_{E}(1,b)/S_{EE}$ as a function of $b$ is similar to  Fig.~\ref{fig:5} for various values of $\alpha$, we only drew the figure of $C_{E}(q,1)/S_{EE}$ and the value range of $\alpha$ in Fig.~\ref{fig:5}  corresponds to the value range of $\alpha$ in Fig.~\ref{fig:4}. From Fig.~\ref{fig:5}, we can see that the capacity of entanglement has a jump when the black hole has a zeroth-order phase transition.

\section{conclusion}
In this paper, we have studied the extended thermodynamics of the hyperbolic AdS black hole with scalar hair by considering the cosmological constant as a thermodynamic variable. The $p$-$V$ phase structure has been shown. The result shows that there are no van der Waals behaviors for the hairy hyperbolic AdS black hole. The phase transition of the two states can exist together was as shown in the $p$-$T$ plane. We also have considered the sign of the heat capacity in the determination of the stability of such black hole.

Moreover, we have extended holographic R\'{e}nyi entropy calculated from hyperbolic black holes with scalar hair. At the fix $q$, the extended R\'{e}nyi entropy  as a function of $b$ which corresponds to changes in the pressure of a black hole. The extended holographic R\'{e}nyi entropy presents a transition at the critical parameters that has been shown when the black hole  has a thermodynamic phase transition at a critical temperature.  Some important limits for extended R\'{e}nyi entropy $S_{q,b}$ from the hairy black hole have been considered. The inequalities of extended holographic R\'{e}nyi entropy are satisfied in our holographic calculations. The conformal dimension of twist operators has also been holographically computed in terms of gravitational quantities.

The capacity of entanglement is known as an important quantum information measure. In this paper, we have defined the extended capacity of entanglement in terms of the extended  R\'{e}nyi entropy. We also show that the extended capacity of entanglement was mapped to the heat capacity of the thermal conformal field theories on the hyperbolic space. It is interesting that the extended holographic capacity of entanglement has a jump when the black hole has a zeroth-order phase transition.

\section*{Acknowledgments}

I thank Xiaoxuan Bai and Jie Ren for helpful discussions. This work was supported in part by the National Natural Science Foundation of China under Grant No. 11905298.

 \bibliographystyle{unsrt}
 \bibliography{reference}
\end{document}